\documentclass[12pt,epsf]{article}
\usepackage[pdftex]{graphicx}
\usepackage{epsfig,amsmath,amssymb,verbatim,mathrsfs,subfigure,feynmf,lpic}
\begin{document}

\def\be{\begin{equation}}
\def\ee{\end{equation}}
\def\bea{\begin{eqnarray}}
\def\eea{\end{eqnarray}}
\def\a{\alpha}
\def\b{\beta}
\def\c{\varepsilon}
\def\d{\delta}
\def\e{\epsilon}
\def\f{\phi}
\def\g{\gamma}
\def\h{\theta}
\def\k{\kappa}
\def\l{\lambda}
\def\m{\mu}
\def\n{\nu}
\def\q{\partial}
\def\r{\rho}
\def\s{\sigma}
\def\t{\tau}
\def\u{\upsilon}
\def\v{\varphi}
\def\w{\omega}
\def\x{\xi}
\def\y{\eta}
\def\z{\zeta}
\def\D{\Delta}
\def\G{\Gamma}
\def\H{\Theta}
\def\L{\Lambda}
\def\F{\Phi}
\def\P{\Psi}
\def\S{\Sigma}

\def\o{\over}
\newcommand{\wt}{\widetilde}
\newcommand{\gsim}{ \mathop{}_{\textstyle \sim}^{\textstyle >} }
\newcommand{\lsim}{ \mathop{}_{\textstyle \sim}^{\textstyle <} }
\newcommand{\vev}[1]{ \langle {#1} \rangle }
\newcommand{\bra}[1]{ \langle {#1} | }
\newcommand{\ket}[1]{ | {#1} \rangle }
\newcommand{\EV}{ {\rm eV} }
\newcommand{\KEV}{ {\rm keV} }
\newcommand{\MEV}{ {\rm MeV} }
\newcommand{\GEV}{ {\rm GeV} }
\newcommand{\TEV}{ {\rm TeV} }
\def\diag{\mathop{\rm diag}\nolimits}
\def\Spin{\mathop{\rm Spin}}
\def\SO{\mathop{\rm SO}}
\def\O{\mathop{\rm O}}
\def\SU{\mathop{\rm SU}}
\def\U{\mathop{\rm U}}
\def\Sp{\mathop{\rm Sp}}
\def\SL{\mathop{\rm SL}}
\def\tr{\mathop{\rm tr}}

\def\IJMP{Int.~J.~Mod.~Phys. }
\def\MPL{Mod.~Phys.~Lett. }
\def\NP{Nucl.~Phys. }
\def\PL{Phys.~Lett. }
\def\PR{Phys.~Rev. }
\def\PRL{Phys.~Rev.~Lett. }
\def\PTP{Prog.~Theor.~Phys. }
\def\ZP{Z.~Phys. }


\baselineskip 0.7cm
\numberwithin{equation}{section}
\begin{titlepage}

\begin{flushright}
IPMU10-0218
\end{flushright}

\vskip 1.35cm
\begin{center}
{\large \bf
Higgs Messengers
}
\vskip 1.2cm
Jason L. Evans, Matthew Sudano, Tsutomu T.~Yanagida\vskip 0.4cm

{\it  Institute for the Physics and Mathematics of the Universe, \\
University of Tokyo, Chiba 277-8583, Japan}

\vskip 1.5cm

\abstract{We explore the consequences of the Higgs fields acting as messengers of supersymmetry breaking. The hidden-sector paradigm in the gauge mediation framework is relaxed by allowing two types of gauge-invariant, renormalizable operators that are typically discarded: direct coupling between the Higgses and supersymmetry breaking singlets, and Higgs-messenger mixing terms.  The most important phenomenological consequence is a flavor-dependent shift in sfermion masses.  This is from a one-loop contribution, which we compute for a general set of weak doublet messengers. We also study a couple of explicit models in detail, finding that precision electroweak constraints can be satisfied with a spectrum significantly different from that of gauge mediation.}

\end{center}
\end{titlepage}

\setcounter{page}{2}

\section{Introduction}

The first rule of supersymmetric model building is that the Standard Model may not know about supersymmetry (SUSY) breaking at tree-level. This is because tree-level splitting of fermions and sfermions tends to give unacceptably light scalars. The second rule is that the communication of the breaking should preserve flavor degeneracy. This is to evade constraints from flavor changing neutral current data. In this work we will violate both of these rules in a natural and viable extension of gauge mediation.

In gauge mediated supersymmetry breaking (GMSB), the hidden sector condition is satisfied by employing a discrete symmetry. Several renormalizable and gauge invariant interactions are forbidden by a $Z_2$ symmetry known as messenger parity. The second rule is then automatically obeyed because the only communication between the SUSY breaking sector and the supersymmetric standard model (SSM) is through gauge interactions, which are flavor blind.

Since this construct is motivated by observational constraints, it is important to understand which aspects of it are strictly necessary and which can be relaxed. In this work we study the consequences of allowing the Higgses to act as messengers themselves, interacting directly with the SUSY breaking sector and mixing with other messengers. Some discussion of the effects of each operator type can be found in the NMSSM literature, for example, and in \cite{raby}. The operators have also been studied in the context of the $\mu/B\mu$ problem in General Gauge Mediation \cite{ggm,muandggm}.

What we will show is that it is possible to turn on both types of interaction such that a novel spectrum is obtained without contradicting current bounds.
Among the more universal aspects of Higgs messenger models are the violation of gaugino mass GUT relations and a light stop.  These follow from the simple facts that the messengers do not form complete GUT multiplets at intermediate energies, and mediation is no longer flavor blind.

\section{Simple Higgs Messenger Model}\label{simple}

We begin by discussing a simple model in which there is a clear distinction between the Higgs and the other messengers. This model will highlight the key features of the more general scenario.  Higgs-messenger mixing will be discussed in sections \ref{general} and \ref{notsosimple}.

Consider an $SU(5)$ GUT model with the typical {\bf 5} and ${\bf\bar5}$ pair of messengers coupled to a SUSY breaking spurion, $\l_SS=M_S+\theta^2F_S$, and allow the MSSM Higgses (triplets already decoupled) to couple to a second spurion, $\l_TT=M_T+\theta^2F_T$.
\be
W\supset \lambda_SS\Phi\wt{\Phi} +\lambda_TTH_uH_d
\ee
Multiple spurions were avoided in \cite{martin} because of the difficulty of suppressing CP violation, but a simple dynamical mechanism for aligning the phases of multiple singlets was recently introduced \cite{esy}.  In fact, the above effective description was obtained such that the $\mu/B\mu$ problem and the CP problem were absent.  Here, however, we will content ourselves with spurions with independent, non-zero scalar and $F$-term vevs. Their function is to split the multiplets, giving the messengers scalars and fermions different masses, $|m_0^2-m_{1/2}^2|\sim F$

Because the Higgs bosons directly feel supersymmetry breaking, they act as messengers and contribute significantly to the sparticle soft masses. These contributions arise at the one-loop level from the ordinary Yukawa interactions.  The first and second generation Yukawa couplings prove to be negligible in this context, but the contributions for the third generation are important.  We will discuss the effects of these interactions later.

\subsection{Flavor Universal Mass Contributions}

Consider first the ${\cal O}(g^2)$ contribution to the gaugino masses and the ${\cal O}(g^4)$ contribution to the sfermion masses, where $g$ is any gauge coupling constant. These calculations are essentially identical to those of minimal gauge mediation \cite{dim, martin}; we just have to sum the contributions from the different messengers.  The one-loop gaugino masses are
\be\label{gmsec2}
M_r=\frac{g_r^2}{16\pi^2}\left(\frac{F_S}{M_S}g(x)+(\delta_{1,r}+\delta_{2,r})\frac{F_T}{M_T}g(y)\right),\qquad x\equiv\frac{\l_SF_S}{M_S^2},\qquad y\equiv\frac{\l_TF_{T}}{M_T^2},
\ee
where $g(t)=t^{-2}(1+t)\ln(1+t)+(t\rightarrow-t)$ and $r=1,2,3$ runs over the Standard Model gauge groups, $U(1)_Y, SU(2)_W, SU(3)_C$ respectively.  The important feature here is the absence of a $T$-sector contribution to the gluino mass, which follows from the decoupling of the Higgs triplets.

The dominant two-loop contribution to the scalar masses has a similar form.
\be
m_{\tilde{f},gauge}^2=2\sum_rC_{\wt f}^r\left(\frac{g_r^2}{16\pi^2}\right)^2 \left(\left|\frac{F_S}{M_S}\right|^2f(x)+(\delta_{1,r}+\delta_{2,r})\left|\frac{F_{T}}{M_T}\right|^2f(y)\right)\label{SferMas}
\ee
where $f(t)=t^{-2}(1+t)\{\ln(1+t)-2\mbox{Li}_2[t/(1+t)]+\frac12\mbox{Li}_2[2t/(1+t)]\}+(t\rightarrow-t)$, and $C_{\wt f}^r$ is the quadratic Casimir for the representation of the $\wt f$ sfermion in the $r$ gauge group.  As with the gauginos, we find a deviation from the ordinary gauge mediation result for the fields charged under $SU(2)_W$ and/or $U(1)_Y$. However, this deviation tends to be small because $F_T/M_T<F_S/M_S$ as we will discuss below.

\subsection{Mass Contributions from Yukawa Interactions}\label{flavor}

The one-loop contribution to the sfermions masses gives the most drastic change to the mass spectrum of ordinary gauge mediation.  Indeed, this model is not purely gauge mediated because there is a direct interaction between messengers (the Higgses) and the quark and lepton multiplets.

\begin{figure}[t!]
\begin{center}$$
\begin{array}{ccc}
\includegraphics[width=1.55in, bb=1.7in 5.34in 7in 7.12in]{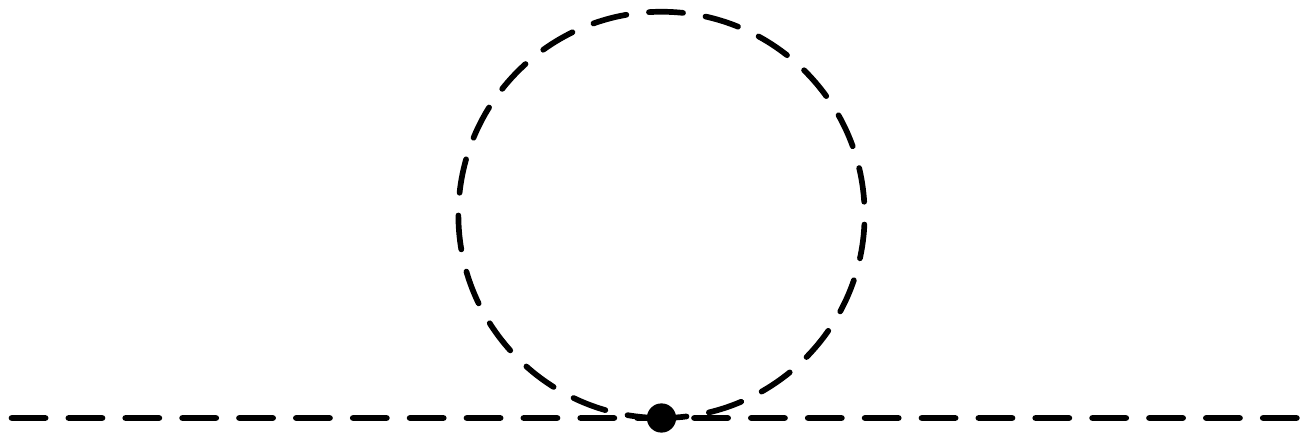}\qquad
&\includegraphics[width=1.45in, bb=1.7in 5.34in 7in 7.12in]{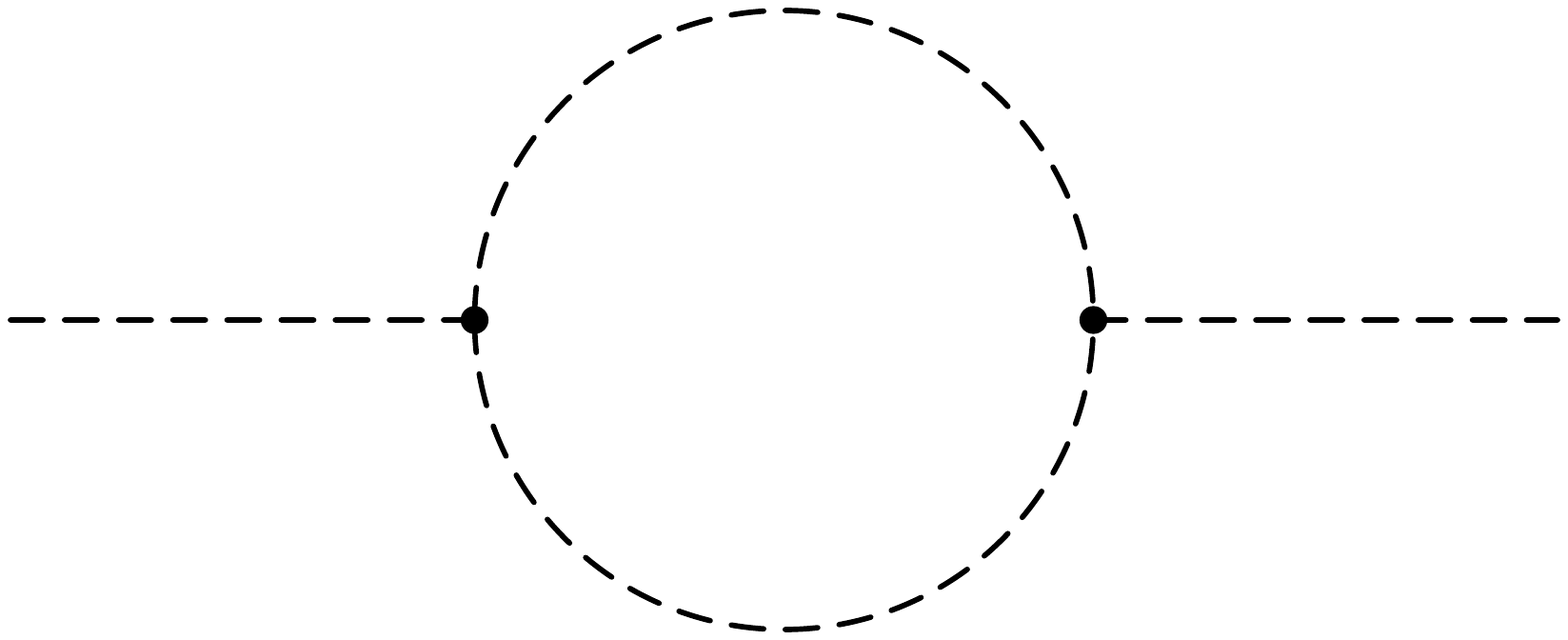}\qquad\quad
&\includegraphics[width=1.45in, bb=1.7in 5.34in 7in 7.12in]{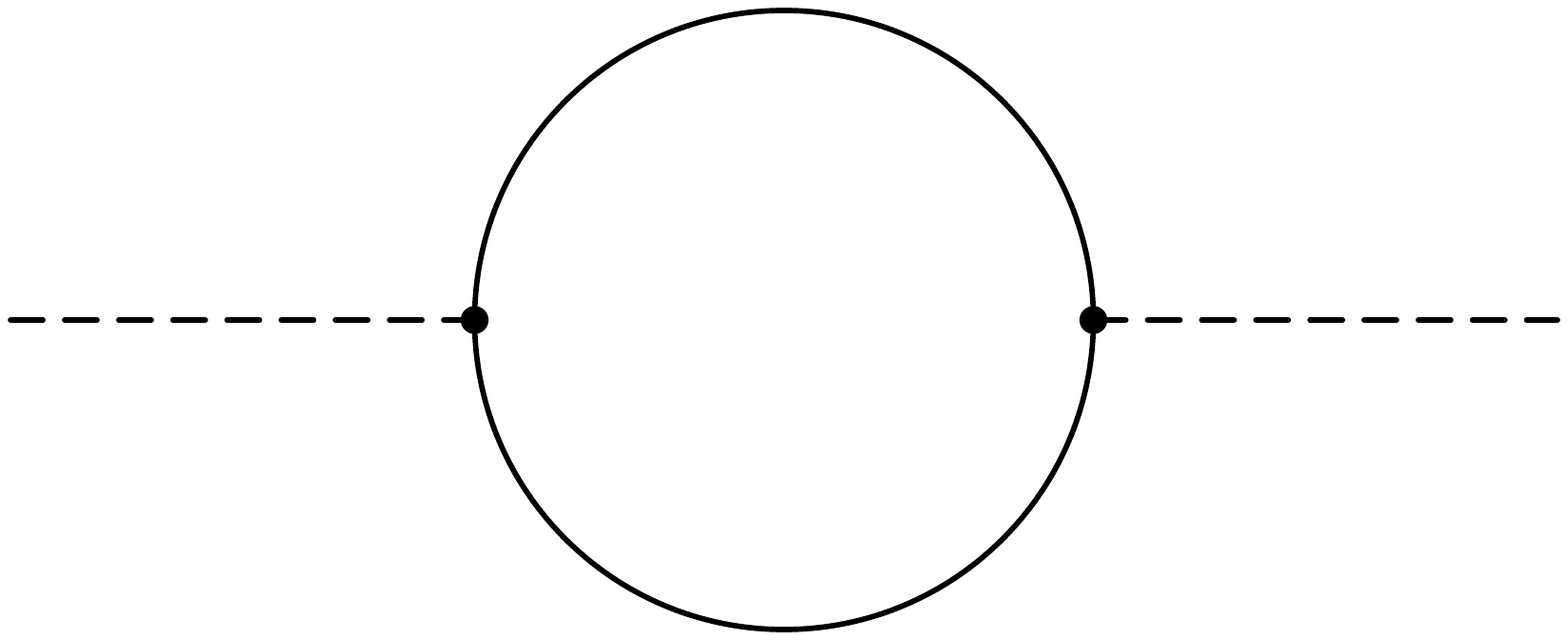}
\end{array}$$
\vskip.1in
\end{center}
\caption{One-loop sfermion soft mass diagrams arising from Yukawa interactions}
\label{1loop}
\end{figure}

The diagrams contributing to the one-loop sfermions masses can be seen in Figure \ref{1loop}. As discussed earlier, we need only consider the contribution to the third generation sfermions. In fact, unless $\tan\beta$ is quite large the only relevant Yukawa coupling is\footnote{Our analysis applies equally well to the large $\tan\beta$ region by replacing stop with sbottom.  However, in that regime the experimental constraints are stronger because of the tight constraints on the $B$ meson mass splitting.} the top Yukawa, $Y_t$. The one-loop contribution to the stop mass is
\bea
m_{\tilde{t},Yukawa}^2
&=&|M_T|^2\frac{|Y_t^2|}{32\pi^2}\left[(2+y)\ln(1+y)+(2-y)\ln(1-y)\right]\nonumber\\
&=& - \frac{|Y_t|^2}{96\pi^2}\left|\frac{F_{T}^4}{M_T^6}\right|+{\cal O}(M_T^2y^6)\label{StoMas}\label{F4}
\label{1loopmod1}
\eea
There are two important things to note about the above expression:
\begin{enumerate}
\item The one-loop sfermion mass contribution is negative semi-definite.
\item The contribution vanishes at leading order in SUSY breaking (${\cal O}(F_T^2/M_T^2)$).
\end{enumerate}
The full stop mass squared, $m_{\tilde{t}}^2=m_{\tilde{t},gauge}^2+m_{\tilde{t},Yukawa}^2$ must be positive to avoid $SU(3)_C$ breaking.  Naively this would seem difficult to accomplish since the negative contribution is ${\cal O}(Y_t^2/16\pi^2)$ and the positive contribution is ${\cal O}((g_3^2/16\pi^2)^2)$.  However, the one loop contribution can easily be suppressed by taking $F_S/\lambda_S M_S >F_{T}/\lambda_T M_T$.  Since the Yukawa contribution vanishes to leading order in SUSY breaking, the hierarchy of these mass scales does not need to be too drastic as long as $y=F_{T}/\l_TM_T^2$ is not close to one.

This makes it possible to have the two-loop contribution dominate.  In the limit of small $x=\lambda_SF_S/M_S^2$ and $y=\lambda_TF_T/M_T^2$, the condition that must be satisfied is
\be\left|\frac{F_{S}}{M_S}\right|^2\left|\frac{M_T}{F_{T}}\right|^2 \left(\frac{g_3^4}{Y_t^2}\right)\left(\frac{f(x)}{\pi^2y^2}\right) >1\label{Two2One}
\ee
Since the couplings are order one as is $f(x)$, we will ignore them. We choose to take $F_{T}\lesssim \lambda_T M_T^2$ $(y\lesssim1)$\footnote{This is contrary to our assumption of small $y$, but a more careful analysis shows this estimate to be adequate.} so that the Higgs mass squared is small and positive and EWSB occurs radiatively.  One could consider tree-level breaking, but this is less convenient computationally and doesn't change the physics significantly. With this constraint, one can see that $F_{S}/M_S$ must be more than a few times larger than $F_T/M_T$.  The deviation from minimal gauge mediation is then ${\cal O}(10\%)$.

\section{General Higgs Messengers}\label{general}

Recently, explicit leading-order (in gauge couplings) mass formulas were calculated for a very general class of gauge mediated models \cite{marques}.  The messenger sector is taken to be
\be
W_\Phi=X_{ij}\Phi_i\Phi_j,\qquad X_{ij}=M_{ij}+\theta^2F_{ij},\qquad i=1,\dots,n
\ee
where $F$ is assumed to be Hermitian.  Using a superfield rotation, $M$ is chosen to be diagonal---its eigenvalues being the messenger fermion masses, $M_i$.  A unitary rotation of the messenger scalars of the form $(\s_1+\s_3)\otimes{\bf1}_{n}$ then brings their mass matrix to the block diagonal form\footnote{We are abusing notation in the usual way by not giving new names to transformed objects.}
\be
m_\Phi^2=\left(\begin{array}{cc}{\cal M}_+&0\\0&{\cal M}_-\end{array}\right),\qquad{\cal M}_\pm=M^2\pm F,
\ee
and the results are given in terms of the matrices that diagonalize these blocks.
\be
m_\pm^2=U_\pm^\dagger{\cal M}_\pm U_\pm=\mbox{diag}(m_{\pm1}^2,\dots m_{\pm n}^2)
\ee
In particular, the gaugino masses are given by \cite{marques}
\be
M_\lambda=\frac{g^2}{8\pi^2}\sum_{i,j,\pm}\pm d_{ij}A_{ij}^\pm M_j\frac{m_{\pm i}^2\ln(m_{\pm i}^2/M_j^2)}{m_{\pm i}^2-M_j^2},\qquad A_{ij}^\pm=(U_\pm^\dagger)_{ij}(U_{\pm})_{ji}
\ee
where $d$ is the Dynkin index normalized so that a flavor of messengers ($\Box+\overline\Box$ of $SU(N)$) gives 1/2.  This formula is insensitive to Yukawa interactions, so it applies equally well to Higgs Messenger models.

Of course, the scalar masses are affected. The ${\cal O}(g^4)$ contribution to sfermion masses is as derived in \cite{marques}
\be
m_{\wt f,gauge}^2\!\!=\!2\!\!\sum_{i,j,r,\pm}\!\!C_{\wt f}^r\left(\!\frac{g_r^2}{16\pi^2}\!\right)^2\!\!\!d_{ij}m_{\pm i}^2\!\left[A_{ij}^\pm\ln\frac{m_{\pm i}^2}{M_j^2}-2A_{ij}^\pm\mbox{Li}_2\!\left(\!1-\frac{M_j^2}{m_{\pm i}^2}\right)\!+\frac12B_{ij}\mbox{Li}_2\!\left(\!1-\frac{m_{\pm j}^2}{m_{\pm i}^2}\right)\!\right]\!,
\ee
where $B_{ij}=(U_\pm^\dagger U_\mp)_{ij}(U_\mp^\dagger U_\pm)_{ji}$.  However, allowing arbitrary Higgs-like couplings to the Standard Model of the form $Y_{iab}\Phi_iQ_au_b\subset W$, there is a one-loop contribution as we saw in the last section.  Of course, many models of this form are already experimentally ruled out.  Here we will just work out the soft masses without considering constraints.  In the next section, we will look in detail at a particularly interesting and viable model.  We will consider the mass for the $u$ squark.  The other sfermions follow without much effort.  The relevant interactions are
\be
-{\cal L}\supset Y_{iab}Y_{icd}^*Q_au_bQ_c^*u_d^*+Y_{iab}Y_{jac}^*\phi_iu_b\phi_j^*u_c^*+(Y_{iab}^*M_{il}\phi_lQ_a^*u_b^*+Y_{iab}\psi_i\psi_{Q,a}u_b+h.c.).
\ee
This is expressed in the basis where $M$ is diagonal, but no other transformations have been made.  Summing the contributions from the diagrams in Figure \ref{1loop}, the mass is found to be
\be
m_{\wt u,Yukawa}^2=-\!\!\sum_{i,j,k,\pm,c}\frac1{32\pi^2}Y_{ica}Y_{jcb}^*\left(4\d_{ij}M_i^2\ln M_i^2-(U_\pm)_{ik}(U_{\pm}^\dagger)_{kj}(M_iM_j+m_{\pm k}^2)\ln m_{\pm k}^2)\right)
\ee
The logs are written with dimensionful arguments to emphasize the fact that the result is UV finite. To recover the previous formula \eqref{1loopmod1}, set $Y_{ica}=Y_t$ and $n=1$, drop all sums except on $\pm$ and take $U_\pm=1$.

\section{Not-So-Simple Higgs Messenger Model}\label{notsosimple}

In this section we will discuss an explicit model where the Higgses and the messengers mix. Before we do that, however, we will briefly digress to discuss Higgs physics.

It was realized long ago \cite{fcnc} that adding Higgs bosons to the Standard Model is not trivial.  Constraints on flavor changing neutral currents prohibit additional Higgs bosons from interacting generically with the MSSM matter content. Proton decay imposes further severe constraints on GUT models.  To address these issues we will impose that
\begin{enumerate}
\item Only one flavor of messengers is permitted to couple to the quarks and leptons.
\item Only these same fields couple to the GUT-breaking adjoint.
\end{enumerate}
These constraints are satisfied in the same way that messengers are sequestered in ordinary gauge mediation---by imposing a discrete symmetry.  Instead of messenger parity, which forbids any direct interaction between the standard model and the hidden sector, the model that we consider has a $Z_3$ symmetry.  This allows us to be consistent with experiment while preserving triplet messengers\footnote{See \cite{raby} for a model in which all messengers are split.}.
We consider the following superpotential.
\be
W=
\lambda_S S \wt H_u\wt H_d +\lambda_T T \left(H_u\wt H_d+ \wt H_uH_d\right)_{SU(2)} \label{SupMix}
\ee
where $H_u$ and $\wt H_u$ are $SU(5)$ fundamentals, and $H_d$ and $\wt H_d$ are anti-fundamentals. The $SU(2)$ subscript on the last term in \eqref{SupMix} indicates that only the weak doublet components of the fields remain because of the decoupling of the Higgs triplets.  $S$ and $T$ are singlets under the gauge group as before:
\be
\l_S S=M_S+F_S\theta^2\qquad \l_T T=M_T+F_{T}\theta^2.
\ee
Under the $Z_3$, the fields transform as
\be\label{z3}
(H,\wt H, S, T)\to (H,\omega \wt H,\omega S,\omega^2 T),\qquad \omega\equiv e^{2\pi i/3}
\ee

For simplicity we have taken the last two terms to have the same coupling.  We will also write our formulas without regard for the phases of our parameters.  As pointed out in \cite{eogm}, for example, one should be concerned about CP violation in models of this sort, but our phases will not remain independent in a complete theory, and it has been shown that even (super)gravitational interactions can sufficiently suppress CP violation in this sort of theory\footnote{In the model of \cite{esy}, cubic terms $S^3$ and $T^3$ were important.  If included, the fields transform under the R-symmetry as $(S,T,H_u,H_d,\wt H_u,\wt H_d,N,{\bf\bar 5},{\bf10})\rightarrow(\frac23S,\frac23T,\frac8{15}H_u,\frac{4}{5}H_d,\frac8{15}\wt H_u,\frac{4}{5}\wt H_d,N,\frac{7}{15}{\bf\bar 5},\frac{11}{15}{\bf 10})$, forbidding several other terms, including the pure-singlet terms like $ST$.}.

\subsection{A Decoupling Limit}\label{decoupling}

To get a feel for this model, we first examine the limit $M_S \gg M_T$. In this limit we can integrate out $\wt H_{u,d}$. The superpotential then becomes
\be
W=-\frac{\lambda_T^2T^2}{\lambda_SS}H_uH_d
\ee
where $S$ and $T$ are superfields. Expanding in powers of $\theta$ we find
\be\label{bmu}
W=-\left[\frac{M_T^2}{M_S}+ \theta^2\left(\frac{M_T^2}{M_S}\right)\left(2\frac{ F_T}{M_T}-\frac{F_S}{M_S}\right)\right]H_uH_d.
\ee
From \eqref{bmu} we see that $\mu$ is small by a see-saw-like mechanism. To get a small $B_{\mu}$, however, we have to tune $2F_T/M_T\approx F_S/M_S$.  With this tuning, a viable electroweak symmetry breaking (EWSB) can be realized. As for the stop mass, with the $\wt H$ integrated out, it is easy to see that the one-loop contribution to its mass will follow exactly as in section \ref{flavor}, giving a negative contribution.  To prevent a tachyonic stop mass in this limit, $B_{\mu}/\mu$ must be smaller than $F_S/M_S$ by more than a factor of a few.  As we will see, away from this limit the model is richer and more interesting.

\subsection{Tree-level Masses and EWSB}

As we move away from the decoupling limit to explore the parameter space, we need to keep at least one light Higgs boson for acceptable EWSB. A light Higgs can only be realized by tuning. This tuning will place us in a region of parameter space where it is difficult, but not impossible, to suppress the one-loop contribution to the stop mass. Because of the additional degrees of freedom in the one-loop contribution to the stop mass, it can actually be made positive for certain regions of parameter space and allows us to see some deviation from the scenario presented above. In the limit presented above, $\mu$ and $B_{\mu}$ were weak scale and therefore produced weak scale higgsinos. Away from this limit there will be regions of parameter space where the higgsinos are much heavier, effectively decoupling from the low-energy theory.

To determine constraints on the parameter space from EWSB, we examine the Higgs bosons masses which are
\be
m_{\mu}^2=M_T^2+\frac{1}{2}(M_S^2+ \zeta_\mu F_S)-(-)^\mu\frac{1}{2} \sqrt{(M_S^2+\zeta_\mu F_S)^2+4(M_SM_T+\zeta_\mu F_T)^2},
\label{HigMas}
\ee
where $\mu=1,2,3,4$ and we have defined $\zeta=(+,+,-,-)$.  By varying $F_S$ and $F_T$, $m_{2}^2$ or $m_{4}^2$ can be made small.  Since the higgsino masses are independent of $F_S$ and $F_T$,
\be
M_{H\a}=\frac{1}{2}\left(M_S-(-1)^\a \sqrt{M_S^2+4M_T^2}\right),\qquad\a=1,2
\ee
they will remain heavy.

For convenience we will take the ``true'' Higgs (the mass-eigenstate responsible for electroweak symmetry breaking) to have a small but positive mass-squared, which then runs negative. This amounts to requiring
\be
M_T^4-F_T^2\gtrsim|M_T(F_SM_T-2F_TM_S)|.
\ee
As we approach the saturation point of this relation, at least one of the Higgses is becoming light.  The expression on the right is proportional to the $B_{\mu}$ parameter in \eqref{bmu}.  When $B_{\mu}$ is tuned to be small this expression will also be small. It was this tuning that allowed us to push the scale of $M_T$ and $F_T$ down without having a tachyonic Higgs. By pushing up the scale of $M_T$, we can relax this tuning.

In this limit, with one light Higgs boson, the low energy theory is very similar to the SM. An important difference is that the Yukawa couplings are larger. Because the electroweak-breaking Higgs boson is a mixture of the four messengers, the interaction of the fermions with the Higgs boson is diluted. The Yukawa coupling must be enhanced to compensate for this suppression. In this model the mixing matrix gives a suppression factor of approximately $1/\sqrt{2}$, so the Yukawa couplings must be enhanced by about $\sqrt{2}$. In all other ways, the low scale Higgs interactions look like the SM.  In fact, the ratio of the Yukawa couplings will be identical to those in the SM. In particular, $Y_t/Y_b\sim 1/40$ and so the sbottom one-loop mass contribution is negligible.  Without this suppression, there would be tight constraints coming from the $B$ meson mass splitting\footnote{ In the limit of vanishing Yukawa couplings for the
1st and 2nd generations and the bottom quark, we can always take a flavor basis where the up-type Yukawa
matrix is diagonal. Thus, there is no FCNC problem.}.

\subsection{Gaugino and Sfermion Masses}

Moving away from the region where $\mu$ and $B_{\mu}$ are EW scale, we have to worry about large negative one-loop contributions to the stop masses. The one-loop stop contribution in this model is much more complicated than in the model considered in section 2. It depends on both $S$ and $T$. This complexity gives us a handle for suppressing this one loop contribution. As stated earlier, in some regions of parameter space the one-loop contribution is positive. To connect with the GUT GMSB limit considered above, we will plot our results in terms of the parameter $z= M_T/M_S$.  Small $z$ corresponds to the decoupling limit of section \ref{decoupling}.

The gaugino masses are given by
\be
M_i=(\frac{3}{5}\delta_{1,r}+\delta_{2,r})g_r^2M_{\chi} +(\frac{2}{5}\delta_{1,r}+\delta_{3,r})\frac{g_r^2}{16\pi^2}\frac{F_{S}}{M_S}g(x)
\ee
where $x=\l_SF_{S}/M_S^2$, $g(x)$ is defined after \eqref{gmsec2}, and $M_\chi$ is calculated and defined in the appendix. The fact that our messengers do not form complete $SU(5)$ multiplets leads to three linearly independent gaugino masses.  This is dramatically different from minimal gauge mediation, where the gauginos only differ by their coupling constants.

\clearpage
\begin{figure}[h!]
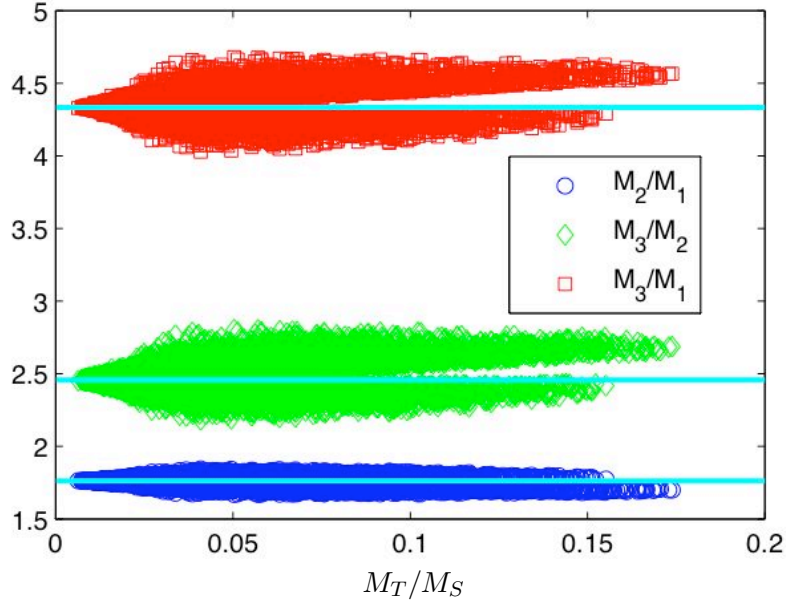

\centering
\begin{lpic}{GauginoRatioFinal(.5)}
\lbl[lr]{113,-7,0;$\mbox{\fontsize{11}{30}\selectfont $M_T/M_S$}$}
\end{lpic}
\vskip.2in
\caption{Ratios of the gaugino masses are shown along with the standard ratios $M_i/M_j=g_i^2/g_j^2$, which are plotted as light blue lines.  Each point has an associated set of parameters for which all mass constraints are satisfied.}
\label{masRat}
\end{figure}
\begin{figure}[h!]
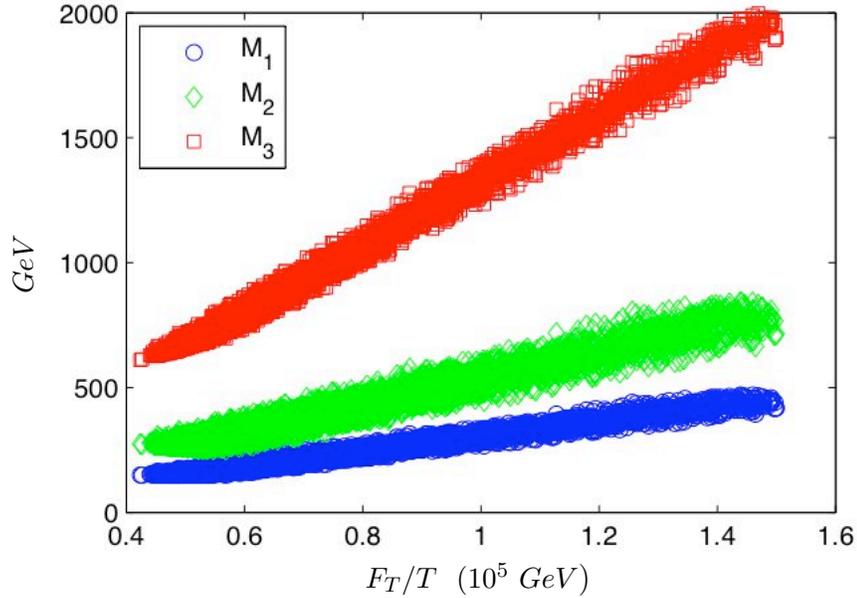

\centering
\begin{lpic}{GauginoMassFinal(.5)}
\lbl[lr]{95,-5,0;$\mbox{\fontsize{11}{30}\selectfont $F_T/T~~(10^5~GeV$})$}
\lbl[lr]{4,71,90;$\mbox{\fontsize{11}{30}\selectfont $GeV$}$}
\end{lpic}
\vskip.2in
\caption{The gaugino masses as a function of one of the SUSY-breaking scales.}
\label{GauMas}
\end{figure}
\clearpage

The violation of the standard GUT relations for the gauginos can be seen in Figure \ref{masRat}, where the ratios of the gaugino masses are plotted. The light blue lines represents the standard GUT mass relations. The gaugino masses do deviate from the
standard GUT relations, but that deviation is limited because of constraints on the stop mass. In this figure, we have required the smallest Higgs mass to be less than a TeV, and imposed a
stop mass bound $m_{\tilde{t}}>176$ GeV.  We applied current experimental bounds \cite{pdg} on all other sparticle masses. If we relax the constraint on the tree-level Higgs mass, the deviation from the gaugino mass relations increases\footnote{We have taken a conservative value for $Y_t$ based on a tree-level analysis. Running effects could allow for a smaller $Y_t$, which would then allow for larger deviations from the GUT gaugino mass relations.}. However, for larger tree-level masses successful EWSB becomes more and more difficult to accomplish\footnote{It is possible to get radiatve EWSB with larger tree-level Higgs masses because of the additional corrections to the Higgs mass in this model.}.

In Figure \ref{GauMas} the gaugino masses are shown.
One can see that the gluino mass in this model is suppressed relative to the standard GUT GMSB scenario for some regions of parameter space. In the standard GUT GMSB the lower bound on the gluino mass is indirectly due to the lower bound on the bino mass. Because the bino mass depends on the function $M_\chi$, its mass may be enhanced relative to the gluino, decreasing the gluino lower bound. However, the one-loop contribution to the stop mass forces us into a region of parameter space where the gluino mass is only mildly suppressed. The gluino mass in this model has to be larger than about $606$ GeV versus $646$ GeV in the standard GMSB scenario.

\begin{figure}[h]
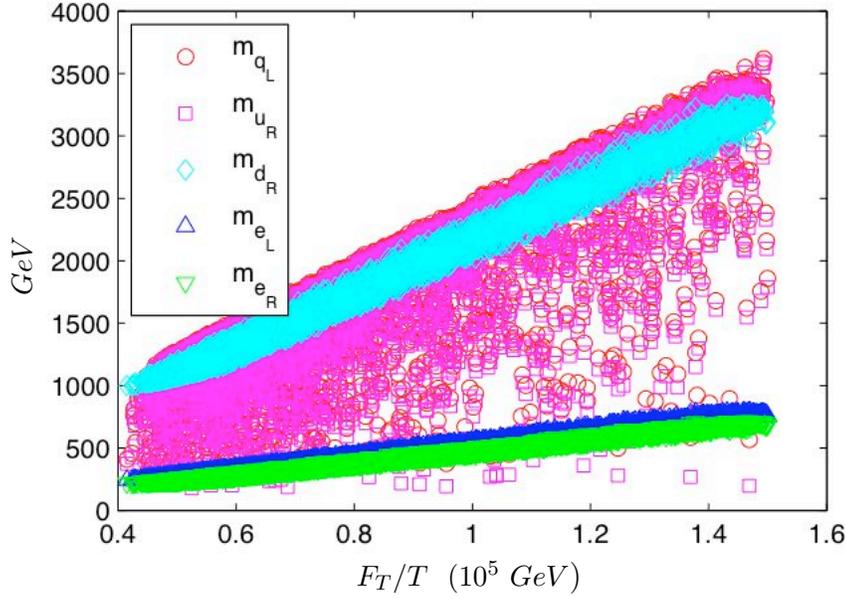

\centering
\begin{lpic}{SfermionMassFinal(.5)}
\lbl[lr]{85,-5,0;$\mbox{\fontsize{11}{30}\selectfont $F_T/T~~(10^5~GeV$})$}
\lbl[lr]{-4,70,90;$\mbox{\fontsize{11}{30}\selectfont $GeV$}$}
\end{lpic}
\vskip.2in
\caption{Sfermion masses are shown as a function of a SUSY breaking scale.}
\label{SfeMas}
\end{figure}

\begin{figure}[h!]
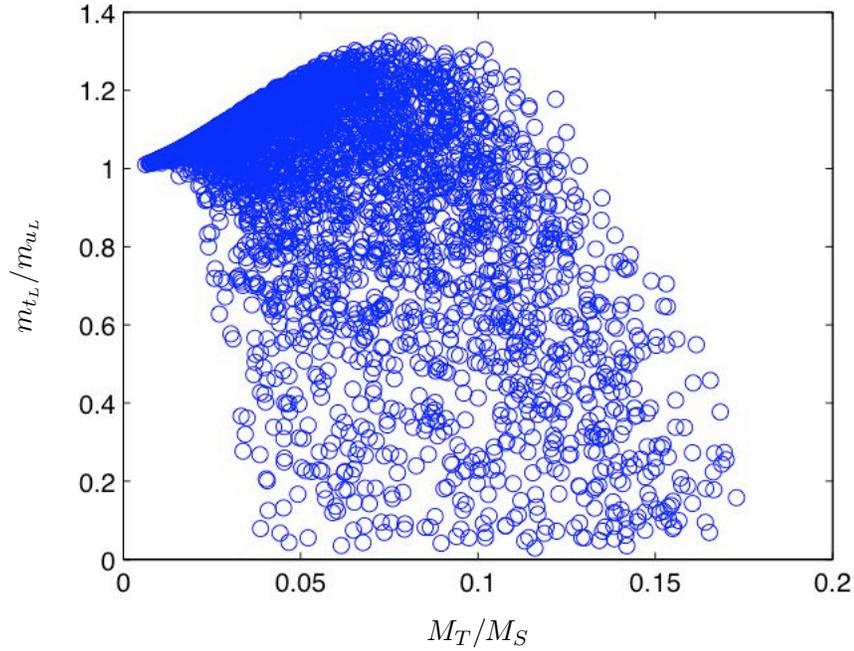

\centering
\begin{lpic}{SquarkRatio(.5)}
\lbl[lr]{112,0,0;$\mbox{\fontsize{11}{30}\selectfont $M_T/M_S$}$}
\lbl[lr]{5,80,90;$\mbox{\fontsize{11}{30}\selectfont $m_{t_L}/m_{u_L}$}$}
\end{lpic}
\vskip.2in
\caption{The stop mass can deviate substantially from the ordinary gauge mediation result as the messenger mass ratio is varied.}
\label{SquRat}
\end{figure}

Next we examine the sfermion masses. They are given by
\be
m_{\tilde{f}}^2\approx\sum_r2C_{\tilde f}^r\left(\frac{g_r}{16\pi^2}\right)^2\!\left[ \left(\frac25\delta_{1,r}+\delta_{3,r}\right)\left|\frac{F_S}{M_S}\right|^2f(x)+\!\left(\frac35\delta_{1,r} +\delta_{2,r}\right){\cal F}\right]\!+\delta_{\tilde{f},\tilde{t}}m_{\wt f,Yukawa}^2
\label{SferMas2}
\ee
where ${\cal F}$ and $m_{\wt f, Yukawa}^2$ are defined in the appendix, $C_{\wt f}^r$ is the quadratic Casimir as before.
The mass spectrum for this model can be seen in Figure (\ref{SfeMas}). The most significant deviation from the standard GUT sfermion masses relations is in the stop. Because the stop has a large one-loop contribution, its mass will be heavily influenced by the size of this one-loop contribution.
To see the deviation in the stop mass from the standard GMSB scenario, we have plotted $m_{u_L}/m_{t_L}$ in Figure \ref{SquRat}.


\section{Final Remarks}

The Higgs sector remains the source of many model-building difficulties, including the infamous gauge hierarchy problem, and, within SUSY, the little hierarchy problem and the $\mu$ problem.  Perhaps this is a clue indicating that the Higgs sector is more complicated than we think.  In this work, we have begun a reevaluation of some of the assumptions associated with the Higgs, and we have found viable non-standard models.

Because the flavor violation is Yukawa-mediated and the Yukawa couplings are small, evading constraints on flavor changing neutral currents is not especially difficult in this scenario.  One might worry about the bottom squark getting a substantial contribution to its mass, but for small $\tan\beta$, the bottom Yukawa coupling is much smaller than the top Yukawa coupling.  So we can still see a deviation in the stop mass while having near degeneracy among the down-type quarks.  This degeneracy is easy to realize to the degree necessary to evade constraints from bottom flavor changing processes.

The other obvious concern in altering the Higgs sector is maintaining successful EWSB. It should be noted that EWSB is quite complicated in these scenarios, but can still be made viable despite outward appearances. In the model of section \ref{notsosimple}, for example, there are additional radiative corrections from $S$ and $T$ loops to aid with radiative EWSB. Furthermore, the quartic term of the single light Higgs boson has additional contributions. These additions, which depend on $\lambda_T$ and $\lambda_S$, lift the tree level mass of the Higgs boson. With the Higgs quartic term supplemented in this way, the upper limit on the tree level Higgs mass is no longer $M_Z$ and large one loop corrections to the Higgs mass from stop loops are unneeded. Thus, distress over light stop masses is unwarranted.

These issues weren't discussed in detail because they can always be tuned away and our goal wasn't a complete and compelling model, but rather a proof of concept. It would be interesting to study this class of models more systematically and thoroughly.  It would be valuable to have a set of necessary conditions on Higgs messenger fields and their couplings. 

\noindent {\bf Acknowledgments:}
This research was supported by World Premier International Research Center Initiative (WPI Initiative), MEXT, Japan.

\appendix

\section{Sparticle Masses}
Here we give some details for calculating the sparticle masses.  One can also obtain these results from the formulas of \cite{marques}. For the reader's convenience, we begin by reproducing the Higgs scalar masses
\bea
m_1^2\!\!\!&=&\!\!\!M_T^2+\frac{1}{2}(M_S^2+F_S)+\frac{1}{2} \sqrt{(M_S^2+F_S)^2+4(M_SM_T+F_T)^2}\nonumber\\
m_2^2\!\!\!&=&\!\!\!M_T^2+\frac{1}{2}(M_S^2+F_S)-\frac{1}{2} \sqrt{(M_S^2+F_S)^2+4(M_SM_T+F_T)^2}\nonumber\\
m_3^2\!\!\!&=&\!\!\!M_T^2+\frac{1}{2}(M_S^2-F_S)+\frac{1}{2} \sqrt{(M_S^2-F_S)^2+4(M_SM_T-F_T)^2}\nonumber\\
m_4^2\!\!\!&=&\!\!\!M_T^2+\frac{1}{2}(M_S^2-F_S)-\frac{1}{2} \sqrt{(M_S^2-F_S)^2+4(M_SM_T-F_T)^2}\nonumber
\eea
and the Higgsino masses
\bea
M_{H1}=\frac12\left(M_S+\sqrt{M_S^2+4M_T^2}\right)\nonumber\\
M_{H2}=\frac12\left(M_S-\sqrt{M_S^2+4M_T^2}\right)\nonumber
\eea

\subsection{Gaugino Masses}

There are four contributions to the generic gaugino---one for each Higgs boson.
Taking our superfields to have the form $H=h+\sqrt2\theta\psi+\dots$, and decomposing ${\bf5}={\bf3}+{\bf2}$ as, for example, $\wt h_u=(\wt h_{u,2},\wt h_{u,3})$ the relevant interactions for this calculation are
\bea
{\cal L_\lambda}
\!\!\!&=&\!\!\!\sqrt{\frac25}g_1 \chi_1^a\Big[
\sqrt{3}(h_u^\dagger\psi_u
+\wt h_{u,2}^\dagger\wt\psi_{u,2}
-\psi_dh_d^\dagger -
\wt\psi_{d,2}\wt h_{d,2}^\dagger)
+\sqrt{2}(
\wt h_{u,3}^\dagger\wt\psi_{u,3}
-\wt\psi_{d,3}\wt h_{d,3}^\dagger)\Big]\nonumber\\
\!\!\!&+&\!\!\!\sqrt{2}g_2 \chi_2^a\tr\Big[T_{SU(2)}^a\big(
\psi_uh_u^\dagger
+\wt\psi_{u,2}\wt h_{u,2}^\dagger-
h_d^\dagger\psi_d
-\wt h_{d,2}^\dagger\wt\psi_{d,2})\Big]\nonumber\\
\!\!\!&+&\!\!\!\sqrt{2}g_3\chi_3^a\tr\Big[T_{SU(3)}^a\big(
\wt\psi_{u,3}\wt h_{u,3}^\dagger
-\wt h_{d,3}^\dagger\wt\psi_{d,3})\Big]
+\mbox{h.c.}\label{GinoInt}\nonumber
\eea
It is convenient to work in the gauge eigenstate basis.  The relevant propagators are found to be
\bea
\vev{h_uh_d} \!\!\!&=&\!\!\!\frac{1}{2}\left[\frac{M_T^2-m_2^2}{(q^2-m_1^2)(q^2-m_2^2)}+
\frac{1}{2}\frac{m_2^2-m_4^2}{(q^2-m_2^2)(q^2-m_4^2)}\right]\!-\!(F_{S,T}\to -F_{S,T})\qquad\label{EvEv}\\
\vev{h_u \wt h_d}\!\!\! &=& \!\!\!\vev{ \wt\phi_u
\phi_d} =\frac{1}{2}\frac{F_T+M_SM_T}{(q^2-m_1^2)(q^2-m_2^2)}\!-\!(F_{S,T}\to -F_{S,T})\label{OdEv}\\
\vev{\wt h_u \wt h_d}\!\!\!&=&\!\!\!\frac{1}{2}\left[\frac{m_1^2-M_T^2}{(q^2-m_1^2)(q^2-m_2^2)}+
\frac{1}{2}\frac{m_2^2-m_4^2}{(q^2-m_2^2)(q^2-m_4^2)}\right]\!-\!(F_{S,T}\to -F_{S,T})\label{OdOd}.
\eea
We have made manifest only one of many symmetries that the propagators possess.  In fact, under exchange of any two of the parameters, $F_S$, $F_T$, $M_S$, and $M_T$, the propagators change at most by a sign.
The gauge eigenstate fermion propagators are
\bea
\langle \psi_u^\a \psi_d^\b \rangle\!\!\!&=&\!\!\!
\frac{M_SM_T^2\e^{\a\b}}{(q^2-M_{H1}^2)(q^2-M_{H2}^2)}\label{PropFe1},\\
\langle \psi_u^\a \wt\psi_{d,2}^\b \rangle\!\!\!&=&\!\!\!\langle \wt\psi_{u,2}^\a \psi_d^\b \rangle
=\frac{ M_T(q^2-M_T^2)\e^{\a\b}}{(q^2-M_{H1}^2)(q^2-M_{H2}^2)},\\
\langle \wt\psi_u^\a \wt\psi_d^\b \rangle\!\!\!&=&\!\!\!
\frac{M_Sq^2\e^{\a\b}}{(q^2-M_{H1}^2)(q^2-M_{H2}^2)}\label{PropFe2},
\eea
Using the above forms of the propagators, the gaugino masses are found to be
\bea
M_1\!\!\!&=&\!\!\!\frac{g_1^2}2\int\frac{d^4q}{(2\pi)^4}\Big[\frac35\left(
\vev{\psi_u \psi_d}\vev{h_u^\dagger h_d^\dagger}
+2\vev{\psi_u\wt\psi_{d,2}}\vev{h_u^\dagger\wt h_{d,2}^\dagger}
+\vev{\wt\psi_{u,2}\wt\psi_{d,2}}\vev{\wt h_{u,2}^\dagger\wt h_{d,2}^\dagger}\right)\nonumber\\
&&\hskip3.23in+\frac25\left(
\vev{\wt\psi_{u,3}\wt\psi_{d,3}}\vev{\wt h_{u,3}^\dagger\wt h_{d,3}^\dagger}\right)
\Big],\nonumber\\
M_2\!\!\!&=&\!\!\!\frac{g_2^2}2\int\frac{d^4q}{(2\pi)^4}\left(
\vev{\psi_u\psi_d}\vev{h_u^\dagger h_d^\dagger}
+2\vev{\psi_u\wt\psi_{d,2}}\vev{h_u^\dagger\wt h_{d,2}^\dagger}
+\vev{\wt\psi_{u,2}\wt\psi_{d,2}}\vev{\wt h_{u,2}^\dagger\wt h_{d,2}^\dagger}
\right),\nonumber\\
M_3\!\!\!&=&\!\!\!\frac{g_3^2}2\int\frac{d^4q}{(2\pi)^4}
\vev{\wt\psi_{u,3}\wt\psi_{d,3}}\vev{\wt h_{u,3}^\dagger\wt h_{d,3}^\dagger}
\label{GauMasFin}.
\eea
The masses can then immediately be evaluated in terms of the following two integrals.
\bea
\G_1(m_1,m_2,m_3,m_4)\!\!\!&=&\!\!\!\int d^4q\frac{1}{(q^2-m_1^2)(q^2-m_2^2)(q^2-m_3^2)(q^2-m_4^2)}\nonumber\\
&&\nonumber\\
\!\!\!&=&\!\!\!\frac{f(m_1,m_3,m_4)-f(m_2,m_3,m_4)}{m_1^2-m_2^2},\\
\G_2(m_1,m_2,m_3,m_4)\!\!\!&=&\!\!\!\int d^4q\frac{q^2}{(q^2-m_1^2)(q^2-m_2^2)(q^2-m_3^2)(q^2-m_4^2)}\nonumber\\
&&\nonumber\\
\!\!\!&=&\!\!\!\frac{m_1^2f(m_1,m_3,m_4)-m_2^2f(m_2,m_3,m_4)}{m_1^2-m_2^2},\\\nonumber
\eea
where
\be
f(m_1,m_2,m_3)=\frac{m_2^2(m_1^2+m_3^2)\ln(m_2^2/m_1^2)-m_3^2(m_1^2+m_2^2)\ln(m_3^2/m_1^2)}{(m_1^2- m_3^2)(m_1^2-m_2^2)(m_3^2-m_2^2)}.
\ee
Combing these results, we find
\bea
\nonumber M_{\chi}=\frac{1}{32\pi^2}
&& \!\!\!\!\!\!\!\!\!\! \Big[M_SM_T^2(m_4^2-m_2^2)\G_1(M_{H1},M_{H2},m_2,m_4)  \\
&& \nonumber \!\!\!\!\!\!\!\!\!\! -M_T^2(M_Sm_2^2+4M_TF_T + 3M_SM_T^2)\G_1(M_{H1},M_{H2},m_1,m_2)\\
&& \!\!\!\!\!\!\!\!\!\! +M_S(m_4^2-m_2^2)\G_2(M_{H1},M_{H2},m_2,m_4)\\
&& \!\!\!\!\!\!\!\!\!\! \nonumber +( M_Sm_1^2 +4M_TF_T + 3M_SM_T^2)\G_2(M_{H1},M_{H2},m_1,m_2) \Big] \\
&& \!\!\!\!\!\!\!\!\!\!\nonumber   -(F_{S,T}\to -F_{S,T}).
\eea

\subsection{Sfermion Mass Calculation}

If we write the scalar mass term as ${\cal L}\supset-{\cal H}^\dagger{\cal M}_0^2{\cal H}$, where
\be
\qquad{\cal H}^T=(h_u,\wt h_u,h_d^*,\wt h_d^*),\qquad
\left(\begin{array}{cccc}
M_S^2&M_SM_T&0&F_S
\\M_SM_T&M_S^2+M_T^2&F_S&F_T
\\0&F_S&M_S^2&M_SM_T
\\F_S&F_T&M_SM_T&M_S^2+M_T^2
\end{array}\right),
\ee
then the rotation to the mass eigenstate basis, $\Phi^\dagger{\cal M}_{diag}^2\Phi$, where ${\cal H}={\cal O}\Phi$, is achieved by the matrix
\be
{\cal O}=\left(\begin{array}{cccc}
A_0 &-A_0'&-B_0&B_0'\\
B_0&-B_0'&A_0&-A_0'\\
A_0&A_0'&-B_0&-B_0'\\
B_0&B_0'&A_0&A_0'
\end{array}\right)\label{HigRot}
\ee
This is related to the $U_\pm$ matrices \cite{marques} (and section \ref{general}) by the rotation that diagonalizes the fermion mass matrices.
Explicitly, we have
\be
A_0^2=\frac{m_1^2M_T^2-m_2^2(F_S+M_S^2+M_T^2)}{2(m_1^4-m_2^4)},\qquad
B_0^2=\frac{m_2^2M_T^2-m_1^2(F_S+M_S^2+M_T^2)}{2(m_2^4-m_1^4)},
\ee
and the primed functions, $A_0'$ and $B_0'$, are gotten by taking $F_{S,T}\rightarrow-F_{S,T}$
we will express our result in terms of the components of this matrix and the masses.
By expanding the gauge eigenstate propagators into a sum of terms with physical mass poles
we can closely follow the calculation of \cite{martin}.  Since the form of the gauge interactions is left unchanged by the basis transformation, little has to be altered.
The interactions that do change are the scalar quartic terms and the gaugino Yukawa interactions.
The $D$-term in the mass eigenstate basis is
\bea
D_H^a= \left( \phi_1^*, \phi_2^*, \phi_3^*, \phi_4^*\right)
\left(
\begin{array}{cccc}
0&0&P&Q\\0&0&-Q&P\\P&-Q&0&0\\Q&P&0&0
\end{array}
\right)T^a
\left(\begin{array}{c}
\phi_1\\\phi_2\\\phi_3\\\phi_4
\end{array}\right)
\eea
where $P=-2(A_0A_0'-B_0B_0')$ and $Q=2(A_0B_0'-A_0'B_0)$ .
And finally, to evaluate the gaugino-mediated diagram, the
chirality-preserving higgsino propagators are needed.  They are
\bea
\nonumber && \langle \psi_u\bar{\psi}_u\rangle=\langle \psi_d\bar{\psi}_d\rangle=A_{1/2}^2\langle \psi_1\bar{\psi}_1\rangle+B_{1/2}^2\langle \psi_2\bar{\psi}_2\rangle\\
&& \langle \wt\psi_u\bar{\wt\psi}_u\rangle=\langle \wt\psi_d\bar{\wt\psi}_d\rangle=B_{1/2}^2\langle \psi_1\bar{\psi}_1\rangle+A_{1/2}^2\langle \psi_2\bar{\psi}_2\rangle\\
\nonumber && \langle \wt\psi_u\bar{\psi}_u\rangle=\langle\wt\psi_d\bar{\psi}_d\rangle=A_{1/2}B_{1/2}(\langle \psi_1\bar{\psi}_1\rangle-\langle \psi_2\bar{\psi}_2\rangle)
\eea
where $A_{1/2}=A_0|_{F_S=F_T=0}$ and likewise for $B_{1/2}$. The corresponding boson propagators are
\begin{eqnarray}
\nonumber &&\langle h_u^*h_u\rangle=\langle h_d^*h_d  \rangle =
A_{0}^2\langle \phi_1 \phi_1^*\rangle +B_{0}^2\langle
\phi_2\phi_2^*\rangle +A_0'^2\langle \phi_3 \phi_3^*\rangle +B_{0}'^2
\langle\phi_4\phi_4^*\rangle\\
&& \langle \widetilde{h}_u^*\widetilde{h}_u \rangle=\langle
\widetilde{h}_d^*\widetilde{h}_d\rangle = B_0^2\langle \phi_1 \phi_1^*
\rangle +A_{0}^2\langle \phi_2\phi_2^*\rangle +B_{0}'^2\langle \phi_3 \phi_3^*\rangle +A_{0}'^2\langle \phi_4 \phi_4^* \rangle\\
\nonumber &&
\langle h_u^*
\widetilde{h}_u \rangle=
\langle h_d^*\widetilde{h}_d \rangle  = A_{0}B_{0}(\langle \phi_1 \phi_1^*\rangle -\langle \phi_2\phi_2^*\rangle) +A_{0}'B_{0}'(\langle \phi_3 \phi_3^*\rangle -\langle \phi_4 \phi_4^*) \rangle
\end{eqnarray}
Applying these changes we find that the two loop contribution to the sfermion masses are given in terms of the function,
\bea
\nonumber {\cal F} &=&\left( \sum_{\mu=1}^4\left( -\langle m_\mu|m_\mu|0\rangle -4m_\mu^2\langle m_\mu|m_\mu|0,0\rangle\right)\right. \\
 &+& \sum_{\a=1}^2\left( -4\langle M_{H\a}|M_{H\a}|0\rangle +8M_{H\a}^2\langle M_{H\a}^2|M_{H\a}^2|0,0\rangle \right) \\
\nonumber &-& 2P^2\left(\langle m_1|m_3|0\rangle +\langle m_2|m_4|0\rangle \right) -2Q^2\left(\langle m_1|m_4|0\rangle +\langle m_2|m_3|0\rangle \right) \\
\nonumber &+& 4\left.\sum_{\mu=1}^4\sum_{j=1}^2 Z_{i,j}\left(\langle m_\mu|M_{H\a}|0\rangle+(m_\mu^2-M_{H\a}^2)\langle m_\mu|M_{H\a}|0,0\rangle\right)\right)
\eea
where
\bea
Z_{ij}=\left(\begin{array}{cc}
2(A_0A_{1/2}+B_0B_{1/2})^2 & 2(B_0A_{1/2}-A_0B_{1/2})^2 \\
2(B_0A_{1/2}-A_0B_{1/2})^2 & 2(A_0A_{1/2}+B_0B_{1/2})^2  \\
2(A'_0A_{1/2}+B'_0B_{1/2})^2 & 2(B'_0A_{1/2}-A'_0B_{1/2})^2 \\
2(B'_0A_{1/2}-A'_0B_{1/2})^2 & 2(A'_0A_{1/2}+B'_0B_{1/2})^2
\end{array}\right)
\eea
The functions, $\langle m_1|m_2|0\rangle$ and $\langle m_1|m_2|0,0\rangle$ are defined in \cite{martin}.

The one-loop calculation follows from the result given in section \ref{general}.
\bea
\nonumber m_{1-{\rm loop}}^2 &=& \frac{|Y|^2}{8\pi^2}\left[ \left(A_0^2m_1^2+B_0^2\lambda^2S^2\right)\ln(m_1) +\left(B_0^2m_2^2+A_0^2\lambda^2S^2\right)\ln(m_2)\right.\\
&+& \left(A_0'^2m_3^2+B_0'^2\lambda^2S^2\right)\ln(m_3) +\left(B_0'^2m_4^2+A_0'^2\lambda^2S^2\right)\ln(m_4) \\
\nonumber  &-& 2A_{1/2}^2M_{H_1}^2\ln\left(M_{H_1}\right)-2B_{1/2}^2M_{H_2}^2\ln\left(m_{H_2}\right)\left.\right]
\eea

\bibliographystyle{distler.bst}
\bibliography{higgsbib}

\end{document}